# Towards 4D modelisation of thermal-field emission from semiconductors

Salvador Barranco Cárceles[1*], Aquila Mavalankar[3]
Veronika Zadin[2], Ian Underwood[1], Andreas Kyritsakis[2]

[1]School of Engineering, University of Edinburgh, Scotland
[2]Institute of Technology, University of Tartu, Estonia
[3]Adaptix Imaging Ltd., Oxford, England
[*]Corresponding author: salvador.barranco-carceles@univ-lyon1.fr, s.barranco.carceles@gmail.com

## ABSTRACT

The theoretical picture of thermal-field emission (TFE) from semiconductors has been limited to 1D and 2D models. This can be attributed to the complex and interdependent phenomena that is involved in TFE from semiconductors which makes the calculations cumbersome. Such limitations result in a partial understanding of the underlying physics of semiconducting surfaces under high electrical fields – which requires the addition of the temporal dimension (4D) to yield a realistic model. Here we develop a 3D model of TFE from semiconductors that can take arbitrary geometries and doping levels. Our model successfully reproduces the characteristic saturation plateau of some semiconductors, as well as its dependence in temperature. The model is found to be in good agreement with experimental data from n-type Germanium at a qualitative level. We propose this model as a platform for future extensions into the full 4D framework, incorporating temporal dynamics for a more complete and predictive description of thermal-field emission from semiconductors.

*Key works:* electron emission, semiconductors, simulations, 3D

## I. Introduction

The unique features of Semiconducting Field Emission (s-FE), such as saturation [1], [2], make it very appealing for highly demanding applications. Additionally, the vast expertise to selectively modify the electronic properties of semiconductors, the ability to machine them accurately, and their low cost put them at an advantage compared with other types of field emitters. However, the underlying physical processes from which their characteristic features arise are complex and difficult to compile in a general modelling and analysis tool.

Several authors [3], [4], [5] established the foundations for the theoretical study of field emission from semiconducting surfaces. However, their models were limited to 1D solutions and did not consider thermal effects or the processes inside the bulk of the emitter. We believe this resulted in theoretical explanations that were not fully satisfactory for the non-linearity of the emitter's I-V curves in Fowler-Nordheim (FN) coordinates. On our previous paper, we explain the physics of s-FE, establish the equations for emission from the conduction and valence bands, set a method to calculate the Nottingham heat, and create a code (GETELEC-2) to evaluate these equations efficiently. However, GETELEC-2 only calculates the emission characteristics, which are the boundary conditions for the processes happening in the emitter. A model attempting to explain the experimental observations made from s-FE requires both accurate calculations of the emitted electrons, resolving the band structure of the semiconductor, and including the thermal effects (Joule and Nottingham). Some work has been done in this direction [6], [7], but it is limited to point or planar emitters and does not take into thermal effects.

Here, we report a general model for field emission from semiconducting surfaces that can take any arbitrary shape or doping profile and include thermal effects. The paper is organised as follows: in section 2, we present and explain the different physical phenomena involved and build the mathematical model. In section 3, we show how the I-V profile of an emitter changes with doping and explain the mechanisms behind emission saturation. In section 4, we compare our results against experimental data and discuss the limits of our model. We finish the paper with conclusions and future research lines.

## II. Model for Field Emission from Semiconductors

The scarcity of available mobile charges is the key differentiating element between a metal and a semiconductor. While in the former, the voltage drops at the surface of the solid, in the latter, the field penetrates the surface and modifies the band structure of the solid. In field emission, with forward bias, the bands bend downward towards lower energy levels as we move towards the field exposed surface. The degree of band bending depends on the applied field and the doping level of the semiconductor, and can be described using the Poisson's Equation [8]:

$$\nabla(-\varepsilon_o \varepsilon_r \nabla V) = \rho \quad (1)$$

where $\varepsilon_o$ is the dielectric permittivity of free space (F/m), $\varepsilon_r$ is the relative permittivity of the semiconductor, $V$ is the applied external potential (V), and $\rho$ is the total space-charge density (C/m$^{-3}$) given by:

$$\rho = q(N_D^+ - N_A^- + n - p) \quad (2)$$

where $q$ is the electronic charge (C), $N^+_D$ and $N^-_A$ are the densities of ionised donors and acceptors (m$^{-3}$) respectively, and $n$ and $p$ are the electron and hole concentrations (m$^{-3}$) respectively as well. Figure 1-a shows a schematic of the system with each of the domains and boundaries, and 1-b depicts the algorithm that our model follows to self-consistently solve all the interdependencies. Our model first solves Eq. 1 (blue loop) for a given applied potential and then evaluates Eq. 2 using the formulae for the respective densities:

$$n = N_C \frac{2}{\sqrt{\pi}} F_{\frac{1}{2}} \left( \frac{E_F - E_C}{k_b T} \right) \quad (3)$$

$$p = N_V \frac{2}{\sqrt{\pi}} F_{\frac{1}{2}} \left( \frac{E_V - E_F}{k_b T} \right) \quad (4)$$

$$N_D^+ = N_D \left[ 1 - \frac{1}{1 + \frac{1}{g} e^{\left( \frac{E_D - E_F}{k_b T} \right)}} \right] \quad (5)$$

$$N_A^- = \frac{N_A}{1 + g\, e^{\left( \frac{E_A - E_F}{k_b T} \right)}} \quad (6)$$

$N_C$ and $N_V$ are the effective density of states in the conduction and valence band, respectively. We have taken them to be constant. The Fermi-Dirac integral $F_{1/2}$ as a function of the Fermi energy is defined as:

$$F_{\frac{1}{2}}(\eta_f) \equiv \int_0^\infty \frac{\eta^{\frac{1}{2}}}{1 + e^{\eta - \eta_f}} d\eta \quad (7)$$

$E_C$, $E_V$, and $E_F$ are the conduction band minimum, the valence band maximum, and the Fermi energies, respectively. $k_b$ is Boltzmann's constant, and $T$ is temperature. $N_D$ and $N_A$ are the donor and acceptor impurities receptively, with $E_D$ and $E_A$ as their ionising energies. $g$ is the ground-state degeneracy factor.

$$E_C = -(V + X_O) \quad (8)$$

$$E_V = -\left( V + X_O + E_g(T) \right) \quad (9)$$

$X_O$ is the electron affinity and $E_g$ is the band gap as a function of temperature:

$$E_g(T) = E_{g0} - \frac{\alpha T^2}{T + \gamma} \quad (10)$$

where $E_{g0}$ is the energy gap at $T = 0$ K and $\alpha$ and $\gamma$ are material specific constants. $E_F$ is determined from the charge neutrality condition:

$$n + N_A^- = p + N_D^+ \quad (11)$$

Now that the electric field and the band structure are known, the emitted electron current from the conduction and valence band are calculated using the expressions from [9], where the details of the equations can be consulted:

$$J_C(E) = f_{FD}(E) \int_{\bar{\alpha}E}^{E} D(E_n) dE_n \quad (12)$$

$$J_V(E) = f_{FD}(E) \int_{\bar{\alpha}E - \alpha E_V}^{E} D(E_n) dE_n \quad (13)$$

Once the current is established in the boundary (Fig. 1-b), from which the continuity condition is applied to calculate the electron current density inside the emitter:

$$\nabla J_n = \nabla J_p = 0 \quad (14)$$

Certainly, since there is current flowing through the emitter, the zero current approximation is not valid, and the positioning of the Fermi level has to take into account the charge distribution and recombination rates:

$$n_1 = \gamma_n n_i e^{\frac{\Delta E t}{k_b T}} \quad (15)$$

$$p_1 = \gamma_p n_i e^{\frac{-\Delta E t}{k_b T}} \quad (16)$$

$$n_i = \sqrt{N_C N_V} e^{\frac{-E_g - \Delta E_g}{2 k_b T}} \quad (17)$$

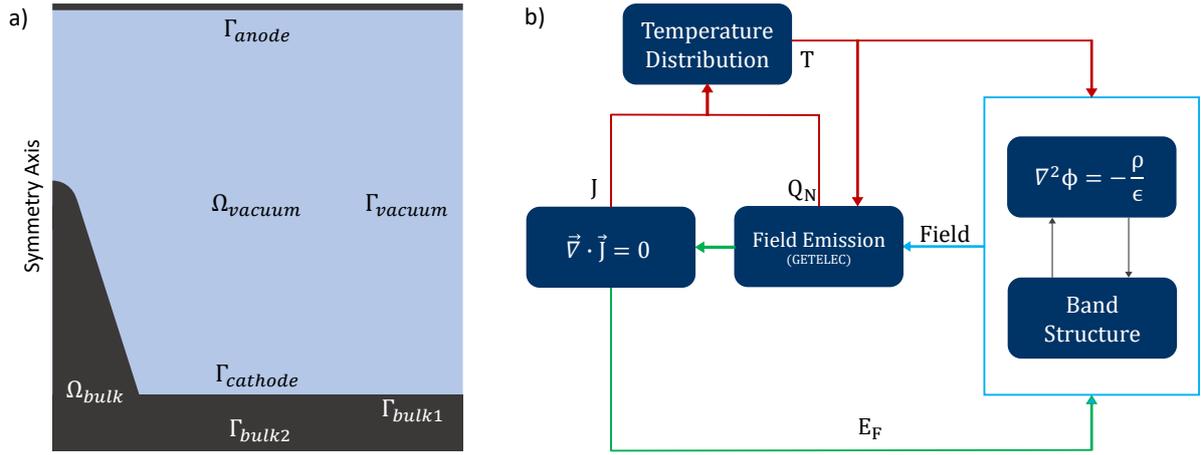

Figure 1: a) domains in model, semiconductor (grey) and vacuum (blue). b) Schematic of the multiple nested effects that are involved in field emission from semiconductors.

$$R_n = \frac{n_p - \gamma_n\gamma_p n_i^2}{\tau_p(n+n_1) + \tau_n(p+p_1)} = R_p \quad (18)$$

$$\nabla J_n = -qR_n \quad (19)$$

$$\nabla J_p = qR_p \quad (20)$$

These new conditions are included in the algorithm, which will find a solution (Fig. 1-b – green loop). Of course, for every iteration of the green loop, the blue one iterates several times until convergence is reached and a new field and band structure are provided.

At this point, the band structure is calculated along the whole emitter and the field emitted current has been determined as well. Now, the Nottingham heat can be calculated as [9]:

$$Q_{N_C}(E) = \int_0^E J_C(E)\,(E-E_R)\,dE \quad (21)$$

$$Q_{N_V}(E) = \int_0^E J_V(E)\,(E_R-E)\,dE \quad (22)$$

$$Q_N = Q_{N_C} - Q_{N_V} \quad (23)$$

where $E_R$ is the replacement energy of the incoming electrons from the bulk. The Joule heating is calculated as:

$$Q_J = \sigma F^2 \quad (24)$$

where $\sigma$ is the conductivity and $F$ is the electric field inside the semiconductor. For high temperatures the radiation loss has to be taken into account:

$$Q_R = -\varepsilon\sigma(T_o^4 - T^4) \quad (25)$$

With these two heat components, the heat transfer equation can be solved:

$$\rho C_P \nabla T + \nabla \mathbf{q} = Q_N + Q_J + Q_R \quad (26)$$

$$q = -k\nabla T \quad (27)$$

$\rho$ is the mass density, and $C_P$ is the specific heat at constant pressure.

As the emitted current rises, the temperature rises as well, increasing the electron and hole concentrations exponentially. The bulk temperature increases also the emission's thermal component [10]. Therefore, all the previous values for the charge and potential distribution, electron emission, and Fermi level have to be recalculated for the new temperature until convergence is met (Fig 1-b – red loop).

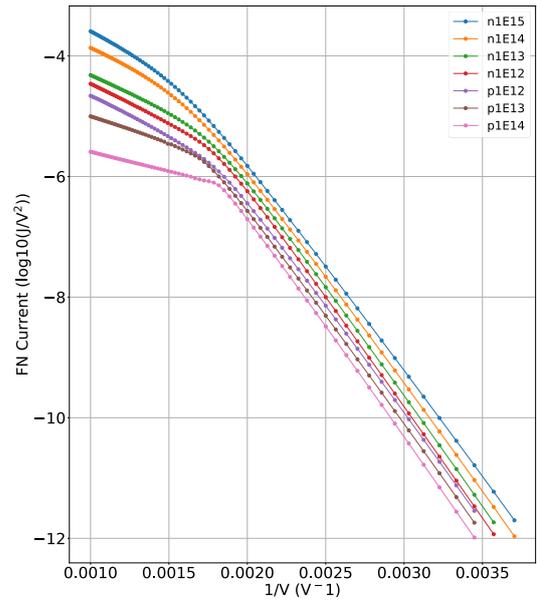

Figure 2: IV curves in FN coordinates for different doping levels. Saturation is present for all doping types, but becoming more evident at higher p-type doping levels.

Now, the system is fully defined and self-consistently solves all the subsystems to provide a complete picture of field emission as a function of electron affinity, band gap, doping levels, temperature, and applied potential. This system is, however, highly exponential and shows several independencies. Thus, to reach a solution, ramping the system up in small steps from $V = 0$ V to the desired final potential is necessary. The system of partial differential equations is solved using the finite element method. Here, we have considered a steady-state solution, and time dependent solutions will be explored in future works.

## III. Field Emission Characteristics of Semiconductors

The main characteristic of emission from semiconductors is the clear deviation from linearity of the current voltage curve plotted in FN coordinates. Any model attempting to simulate the behaviour of s-FE must be able to reproduce this non-linear FN plot and provide a physical explanation for the apparent saturation. Our model uses as inputs an arbitrary geometry. For the purpose of this manuscript an axisymmetric conical shape has been used input to mimic the shape of a tip like emitter. Due to the availability of experimental data from germanium, we have chosen this element to run the simulations and show the emission properties of the emitter for a broad range of parameters (e.g., fields, impurity levels, and temperatures).

Figure 2 shows I-V curves in FN coordinates for the same geometry, at a starting temperature $T = 293$ K, and different doping levels from n-type to p-type. For all cases, the resulting IV curves show a quasi-linear behaviour in the low current Region I, followed by the saturation regime of Region II. This characteristic is observed in all calculations regardless of the doping levels but becomes more pronounced as the acceptor impurity level rises. The emission characteristic is essentially the same for doping levels below $D < 1 \cdot 10^{11}$ cm$^{-3}$. Saturation is reproduced and observed for all doping levels and doping types, but the effect becomes stronger as the acceptor impurity levels are higher.

From a field emission experiment, three parameters are commonly extracted: emission area $A_{fe}$, field enhancement factor $\beta$, and work function $\phi$. Our model allows to study the dynamics of these parameters as function of the emitter geometry, doping, applied potential, and temperature. First, we study how the emission area evolves with the applied potential. We define the emission area using the concept of *notional area* defined by:

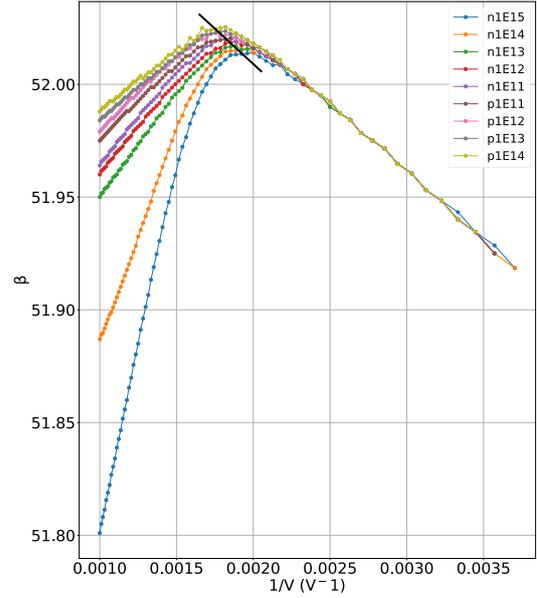

Figure 3: Field enhancement factor for a Ge field emitter as function of the inverse of the potential and doping.

$$A_r = \frac{I}{j_r} \quad (28)$$

where $A_r$ is the notional area (nm$^2$), $I$ is the total emitted current (A) and $j_r$ is the maximum current density (A/nm$^2$). The notional area grows with an increasing potential (Fig. 3) and it has a much lower value than the apparent macroscopic emission area ($A_M$ = 1134 nm$^2$ for the simulated emitter) – the macroscopic emission area is defined by calculating the area using the observable radius of the tip. This is in agreement with previously reported values [11]. The emission area seems to grow linearly with the tip radii (Fig. 4) The cone angle also influences the emission area, while the emitter height and the doping level or type do not seem to have an effect on the emission area - their trends have not been plotted for brevity.

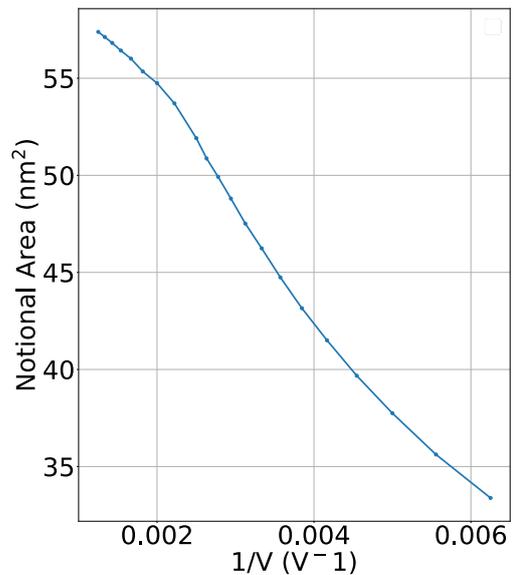

Figure 4: Increasing emission area of a Ge emitter with rising potentials.

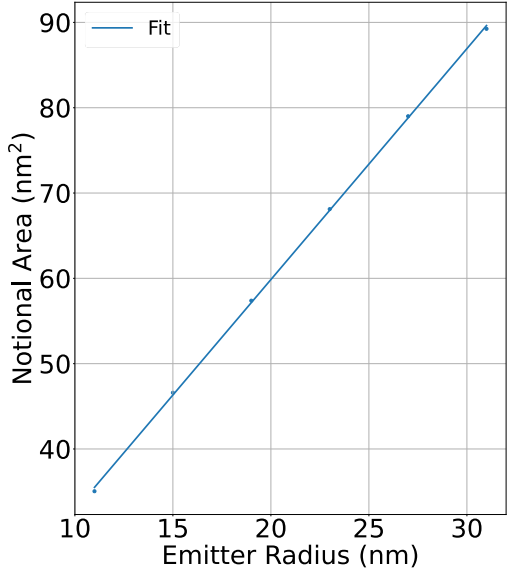

Figure 6: Emission area as function of the tip radius.

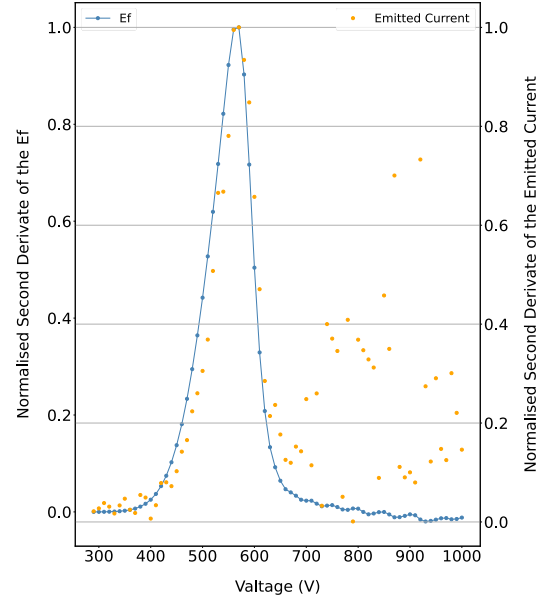

Figure 5: Second derivates of Ef and I, where it can be seen that the change of current linearity coincides with a change of a change of Ef.

The field enhancement factor β is another relevant factor in field emission and it has been proposed to play a determinant role in the appearance of Region II – pattern shrinking process [12]. We have calculated, for several doping levels and as a function of the potential, as follows:

$$\beta = \frac{E_{max}}{E_{macro}} \quad (29)$$

where $E_{max}$ is the field at the apex of the emitter (V/m), and $E_{macro}$ is the macroscopic field that results from applying a potential $V$ between the anode and cathode (Fig. 2.10):

$$E_{macro} = \frac{V}{d} \quad (30)$$

Figure 5 shows that for semiconductors β varies with the applied potential, with the same rising rate from low fields to a point at which the trend is reversed, and differences between doings emerge. The inflexion point ($V$ ~580 V) coincides with the point at which saturation occurs (Fig. 2), so saturation could be attributed to a change in enhancement factor, but two observations rule otherwise. First, the drop in field enhancement factor could be considered negligible due to the small variance between the maximum value and its minimum – a drop of ~0.5% cannot be accounted for saturation over a range of 300 V where the field almost doubles in value. Secondly, the β drop is more pronounced on n-type samples, which exhibit the weakest saturation. Thus, the role of the beta enhancement factor on the saturation of the I-V curve can be ruled out.

Since the drop of the field enhancement factor cannot explain the current saturation, the question of its origin should answer to another physical process. The *retarded* inflexion point of β as the doping becomes more p-type provides hint of to find the origin of current saturation (black line in Fig. 5). The slope of the black line is the same as that of β on the right-hand side region of the graph, which coincides with the metallic behaviour of the s-FE or Region I. The metallic region results from a constant Fermi level, i.e., a regime where the zero current approximation can be applied. When the charge supply cannot keep up with the current drawn from the emitter, there is a potential drop (i.e., a quasi-Fermi level gradient) inside the semiconductor – the point from which the emission loses its metallic-like properties and becomes characteristically semiconducting-like. If this analysis is to be true, the point at which saturation starts occurring must be the point at which the position of the Fermi level changes. Figure 6 shows that the saturation regime coincides with a of drop of the fermi level, which is consistent with previously reported mechanisms for current saturation [6].

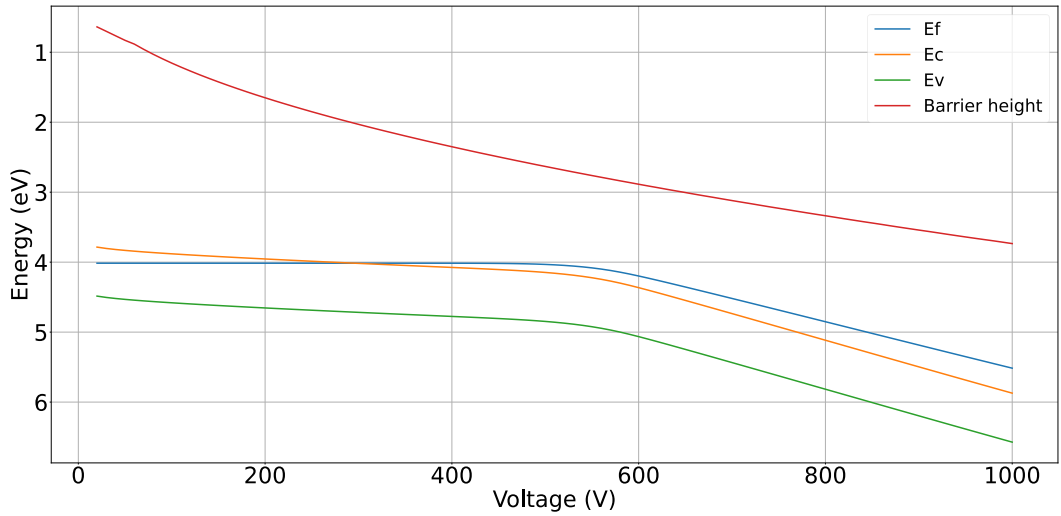

Figure 7: Potential barrier height and Fermi level as a function of the applied potential.

To illustrate the mechanism responsible for saturation in semiconductors, we have plotted the band structure and the height of the potential barrier as a function of the applied voltage (Figure 7). At low voltages ($V < 200$ V, for our simulations) there is effectively no emission because the top of the potential barrier is high with respec of $E_C$ (~2.7 eV). As the voltage is increased the potential barrier is lowered and $E_C$ falls below $E_F$ ($V \cong 400$ V), which results in an exponential *metal-like* growth of the emitted current. The exponential growth of the emitted current continues until the lack of available charges for conduction results in a drop of $E_F$ ($V \cong 600$ V). From this point, the distance between the top of the potential barrier and $E_F$ remains roughly constant with a value ~ 1 eV. The conduction and valence bands follow the same trend. The fact that a lowering of the potential barrier is followed by a lowering of the bands results in a regime where the electrons experience an approximatelly constant transmission coefficient; thus explaining the physical origing of the saturation regime.

It is well documented that the I-V curve from a s-FE, when in the saturation regime, is very sensitive to temperature [1] and light [12]. While the physics for photo enhanced field emission goes beyond the scope of the present work, the effect of temperature change can be readily studied for different doping levels. Figure 8 shows the resulting I-V curves in FN coordinates for the p1E13 emitter,

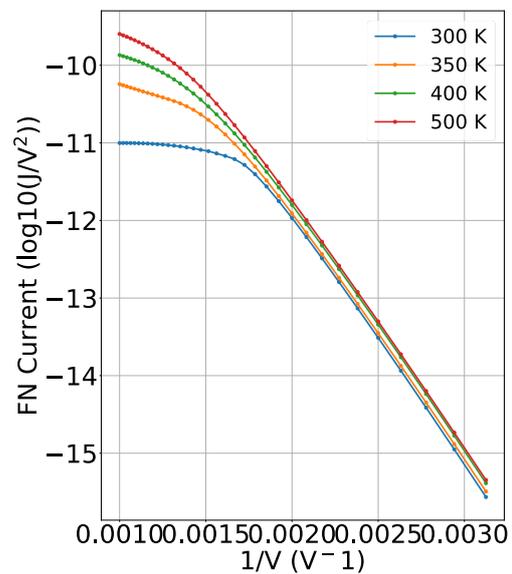

Figure 8: IV curve of a saturated emitter in FN coordinates as a function an applied external temperature.

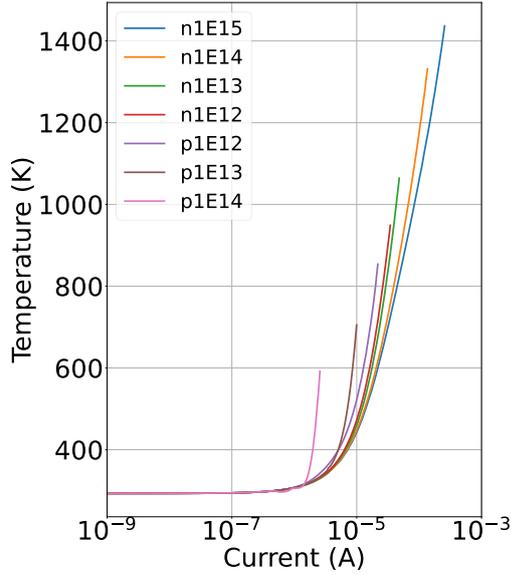

Figure 9: Temperature rise as a function of emitted current for different doping levels.

and it can be observed that our model is able to replicate reported experimental results.

The fact that the saturation regime is so sensitive to temperature further reinforces the explanation of voltage drop in the semiconductor, as the carrier supply is exhausted. The intrinsic carrier density of a semiconductor is very small compared to doping levels. However, the intrinsic carrier density increases very rapidly with temperature, approximately doubling every 15 K for germanium. This means that for rising temperatures more carriers become available for conduction and the Fermi level *relaxes,* bouncing back to its original position. This results in an exponential increase of the emitted current, until more current is drawn from the emitter than that that can be supplied, and saturation appears.

A field emission model cannot be considered completed if the thermal effects are not included, since temperature is such an important parameter in field emission because it draws a hard limit on the current that an emitter can yield. Additionally, physic/chemisorption dynamics depend on it and can affect the emission itself (see previous discussion). Being informed about the likely temperature of an emitter is also useful since it requires a great experimental effort and specialised equipment to estimate the temperature by looking at the Fermi tail of an electron energy distribution. Thus, our model takes into account the Joule and Nottingham heat components to calculate the emitter's temperature.

Figure 9 shows the maximum emitter temperature as a function of the total emitted current from emitters with different doping levels. The graph suggests that the more p-type the emitter becomes, the higher the temperature for the same emitted current. Such orderly trend is also an unexpected result. In general terms, the thermal conductivity is proportional to the mean free path of the phonons. In a perfect crystal, since there are not scattering points, the mean free path tends to infinity and thus it does the thermal conductivity. Therefore, for a given current, low doping level samples are expected to suffer a lower temperature rise than high doping level samples, since the later present a higher density of scattering points that shortens the mean free path of phonons. However, the arising trend from our simulations depicts another behaviour. Currently, because of the very complex nature of thermal effects (e.g.; differences in the life time of electrons and holes, in their mass, in their velocity,…) we do not have a wholistic and satisfactory explanation of the results show on Fig. 9 nor experimental data against which compare our results. Our hypothesis is that the voltage drop plays a key role since the dissipated power is:

$$W = \Delta V \cdot I \qquad (31)$$

where $W$ is the power (W), $\Delta V$ is the potential drop (V), and $I$ the current through the emitter's shank. Because the voltage drop is more pronounced on highly doped p-type semiconductors, more energy is dissipated on them that that compared with the same doping level for n-type. Then, as $E_F$ drops more power is dissipated, which increases the temperature, which reduces the thermal coefficient, which increases the temperature, … This analysis needs to be put in context of scattering and velocity drift saturation that occurs in semiconductors at high fields and high temperatures.

## IV. Experimental Validation of the Model

Field emission, without the complexity of semiconductors, is an already hard phenomenon to model. Besides, we have made a series of assumptions and simplifications that might compromise the validity of our model. To verify the validity of our model, we have simulated a germanium emitter like that reported by Shepherd and Peria (S&P) [13]. We chose their work because the provide information about the emitted current, the fields at the surface of the emitter, electron energy distributions, and the variations of the Fermi level. Equally importantly, their experiments were performed in ultra-high vacuum, with well-conditioned emitters to the point that clear field emission patterns were obtained. Besides, the geometrical characterisation of their germanium emitters is reported in [14], from which we can extract the dimensions of the emitter to create a faithful simulation of their system.

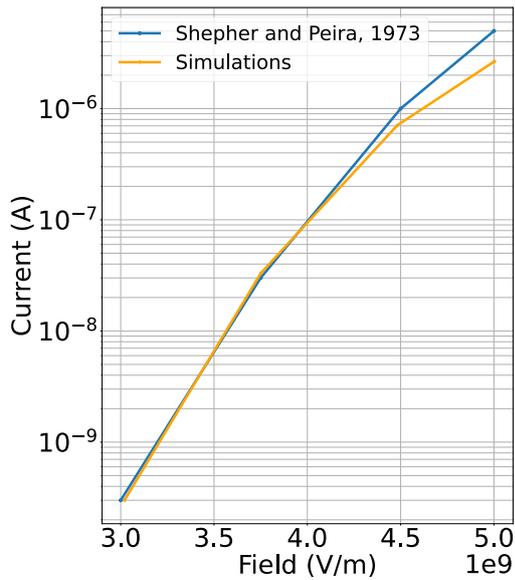

Figure 10: Comparison between experimental data [85] and our simulations of field emitted current as function of the electric field.

The simulation parameters (e.g., tip radius, emitter angle, and doping level) are extracted from [13], [14]. The height of the emitter is not reported, so we have selected a height obtained from the images of their manuscript. Because we are uncertain about the height of their emitters and the cathode to anode distance, we cannot compare the I-V curve. However, they reported the field values that correspond to the total current they measured (Fig. 6 [13]). Therefore, we have consulted the simulated current that correspond to the field values that they reported. Figure 10 shows a comparison between the experimental and simulated I-F (current-field) and it can be seen that they are in good agreement. Over four orders of magnitude the maximum discrepancy is a factor of two, were our simulation yields a lower current than the reported one. In an attempt to explain the discrepancy in the high current regime, we propose that it is likely the result of the contribution of the surface states to the emission - that our simulation does not include. It is also about the nature of the surface states emission and the bulk emission, and in light of the results that we have obtained, that we disagree with S&P about the interpretation of their results.

Figure 11-a shows one of the experimental electron energy distributions from germanium from [13]. On the graph, one can see a larger peak (that presents a stable shape over the range of fields reported) and a smaller peak (that grows in magnitude as the applied field grows). S&P attribute the former to electrons coming from the surface states and the

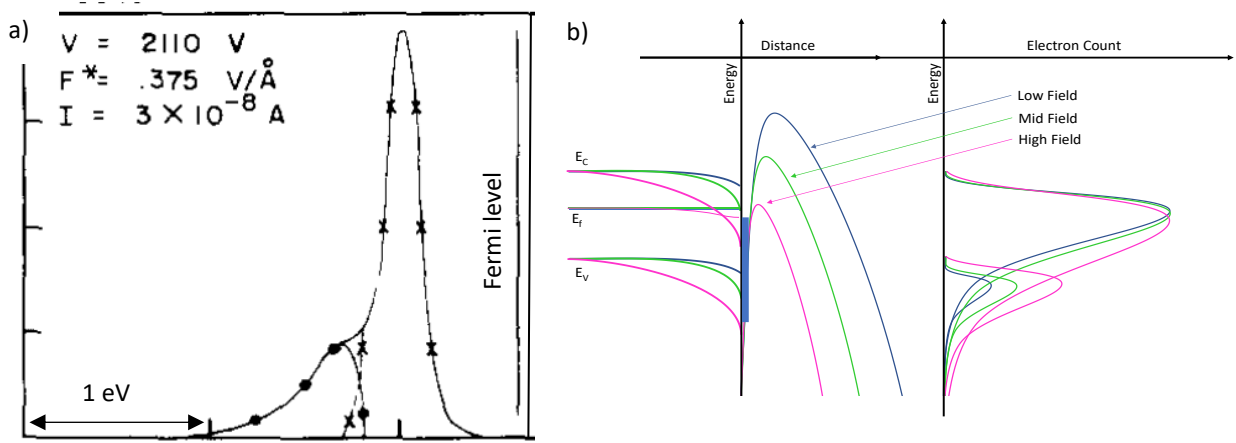

Figure 11: a) Electron energy spectrum extracted from [84]. B) Representation of the band structure, potential barrier, surface states (blue bar), and electron energy distribution for three fields strengths.

later to electrons coming from the valence band. We propose an alternative interpretation of their results based on two observations from our model: firstly, while field emission from the valence band is possible, its contribution to the total emitted current is negligible compared to that from the conduction band in a regime where the current is measurable (Fig. 6 in [9]). Secondly, if the smaller peak is to be attributed to electrons from the valence band, then, there should be a conduction band peak 0.7 eV *to the right* of the valence band peak and the band gap should be visible on the spectrum. We argue that the large peak is to be attributed to the conduction band and the small peak to the surface states of clean germanium.

Our alternative explanation assumes that the native surface states of clean germanium are somewhat below the Fermi level. Because of their low energy, at low and mid fields their contribution to the total emitted current is small due to the lower transmission they experience (Figure 11-b, blue line). As the field increases their contribution grows due to the fact that the potential barrier becomes thinner. However, the emission is led by the conduction band because it is not yet saturated and has a higher transmission coefficient (Figure 11-b, green line). At higher fields, the Fermi level drops resulting in saturation of emission from the bulk, but the contribution from the surface states grows as the potential barrier becomes thinner and thinner (Figure 2.20, pink line). Based on the results from our model, we believe that this process satisfactorily explains the observations by S&P. However, the matter remains open to alternative explanations until more theoretical (e.g., calculation of the germanium surface states by DFT) and experimental work (e.g., energy spectrum over a broader range of fields) are done to study the dynamics of surface states in semiconductors.

We suggest a drop of the Fermi level to explain the formation of the saturation region, and invoke the same effect to propose an alternative explanation that better reconciles our model with experimental observations. The drop of the Fermi level should result in a shift of the position of the electron spectrum, as these are measured for increasing fields. S&P observed the drop of the Fermi level on a retarding potential electron analyser (Fig 12 orange lines) and our simulations (Fig 12 blue lines) qualitatively agree with their measurements. While the magnitude of the drop is not well reproduced, the drop occurs at the same field values.

## V. Conclusions

Semiconducting field emitters offer interesting features for scientific studies and applications. Despite their potential, their utilization is hindered because of their complexity and the fact that they are not fully understood, so the design and specification processes cannot be successfully implemented.

This paper gathers our efforts to develop a general thermo-field emission model for semiconductors, upon which more advanced physics can be implemented (e.g.; quantum confinement or photofield emission). The present model produces the characteristic non-linear I-V curves of semiconductors and its temperature dependence for arbitrary geometries and doping levels. The model provides a 3D picture of the of the distribution of charges, field, band structure, and temperature of an emitting semiconductor under high fields. In so, it has allowed us to have a clearer picture of the mechanism behind the saturation regime and exponential growth emission in Region II with increasing temperatures. A comparison of the model with experimental data is in good agreement, at least qualitatively. We acknowledge that quantitative comparison is not yet possible due to the assumptions and simplifications done to build this model (e.g.; free electron model and no surface states).

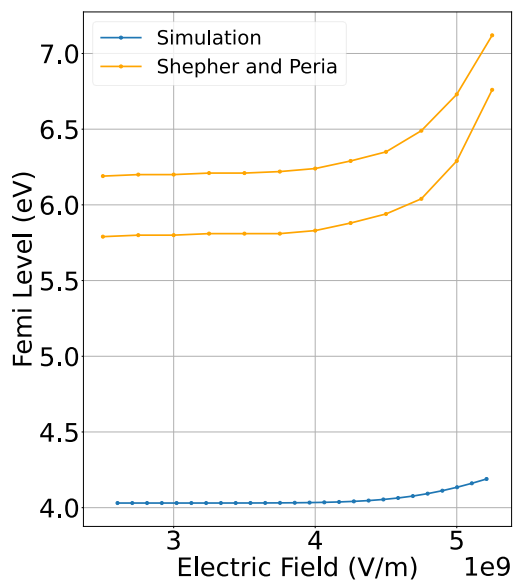

Figure 12: Fermi level drop as a function of the applied field: experimental (orange) [13] and simulation (blue) values.

For the future and more sophisticated version, we are already working in adding the temporal dimension to study time related phenomena like electron transport and heat dissipation. We hope we can encourage the community to build upon this model to include the physics of photon enhanced field emission, quantum confinement, and surface states.

*ACKOWLEDGEMENTS*
S.B.C acknowledges support from the Royal Society of Edinburgh (Saltire Fellowship Grant No. 1956). S.B.C. also acknowledges Dr. Anthony Ayari and Dr. Stephen Purcell for the engaging and constructive discussions held during this work.

*AUTHOR DECLARATIONS*
**No conflict of interest.** The authors have no conflicts of interest to disclose.